\documentclass[twocolumn,preprintnumbers,amsmath,amssymb]{revtex4-1}
\usepackage{graphicx}
\usepackage{dcolumn}
\usepackage{bm}

\begin{document}

\title{Toward the creation of terahertz graphene  injection laser }
\author{
V. Ryzhii$^{1}$, M. Ryzhii$^{1}$, V.~Mitin$^2$, and T. Otsuji$^{3}$} 
\address{
$^1$ Computational Nanoelectronics  Laboratory, University of Aizu,
Aizu-Wakamatsu 965-8580, Japan}
\address
{$^2$ Department of Electrical Engineering, University at Buffalo,
Buffalo, NY 1460-1920, USA
}
\address
{$^3$ Research Institute for Electrical Communication, Tohoku University, Sendai 980-8577, 
 Japan} 

\begin{abstract}
We study the effect of population inversion associated with the electron and hole injection in graphene p-i-n structures at the room and slightly lower
temperatures. 
It is assumed that the  recombination and energy relaxation of electrons and holes
is associated primarily with  the interband and intraband processes assisted by
  optical phonons. 
The dependences of the electron-hole and optical phonon effective temperatures
on the applied voltage, the current-voltage characteristics,
and the frequency-dependent dynamic conductivity are calculated.
In particular, we demonstrate  that  at low and moderate voltages
the injection can lead to a pronounced cooling
of the electron-hole plasma in the device i-section  to the temperatures below the lattice temperature.
However at higher voltages, the voltage dependences can be ambiguous
exhibiting the S-shape.
 It is shown that
 the frequency-dependent dynamic conductivity  can be negative
in the terahertz range of freqiencies at  certain values of the applied voltage.
The electron-hole plasma cooling  substantially reinforces the effect of negative dynamic conductivity and promotes the realization of terahertz lasing. On the other hand, the heating of optical phonon system
can also be crucial affecting the realization of negative dynamic conductivity
and terahertz lasing at the room temperatures. 
\end{abstract}
\maketitle
\newpage
\section{Introduction}

The gapless energy spectrum of electrons and holes 
in graphene layers (GLs), graphene bilayers (GBLs),
 and non-Bernal stacked multiple graphene layers (MGLs)~\cite{1,2,3}, opens up prospects of creating terahertz (THz)  lasers based on these
graphene  structures.
In such structures, GLs and MGLs with optical
~\cite{4,5,6,7,8,9,10} and injection~\cite{11} pumping can exhibit 
the interband population inversion and  negative dynamic conductivity 
in the THz range of frequencies and, hence,
can serve as active media in THz lasers. The most direct way to create 
the interband population in GLs and MGLs is to use optical pumping~\cite{4}
 with the photon
energy $\hbar\Omega_0$ corresponding to middle- and near- infrared (IR) ranges.
In this case, the  electrons and holes, photogenerated with the kinetic energy
$\hbar\Omega_0/2$, transfer their energy to optical phonons and concentrate
in the states near the Dirac point~\cite{4,12,13}.
The amplification of THz radiation  from optically-pumped 
GL structures observed 
recently~\cite{14,15} is attributed to the interband stimulated emission.
However, the optical pumping with relatively high photon energies exhibits
drawbacks. First of all, the optical pumping, which requires complex 
setups,  might be inconvenient method
in different applications of the prospective graphene THz lasers.
Second, the excessive energy being received by the photogenerated electro-hole plasma from pumping source can lead to its marked heating because of
 the redistribution
of the initial electron and hole energy $\hbar\Omega_0/2$
among all carries due to rather effective inter-carrier collisions.
The latter results in a decrease of the ratio of the quasi-Fermi energies $\mu_e$
and $\mu_h$ to
the electron-hole effective temperature $T$ 
that, in turn, complicates achieving of sufficiently large values
of the dynamic conductivity. As demonstrated recently~\cite{16,17},
the negative conductivity at the THz frequencies  is very sensitive
to the ratio of the photon energy $\hbar\Omega$ and the optical 
phonon energy $\hbar\omega_0$, as well as to the relative efficiency 
of the inter-carrier scattering and the carrier scattering on optical phonons.
The decay of nonequilibrium optical phonons also plays an important  role.

The abovementioned complications can be eliminated in the case
of pumping resulting in the generation in GLs electrons and holes with
relatively low initial energies.
This in part  can be realized in the case of optical pumping with 
$\hbar\Omega_0/2 < \hbar\omega_0$`\cite{16} . Taking into account that in GLs
$\hbar\omega_0 \simeq 0.2$~eV, in the case of CO$_2$ laser as a pumping source, 
$\Omega_0/\omega_0 \simeq 0.5$. As shown~\cite{16}, in such a case,
the electron-hole plasma can even be cooled, so that $T < T_0$, where $T_0$ is the lattice (thermostat) temperature.
Another weakly heating or even cooling
 pumping method which can provide low effective temperature $T$ (including $T < T_0$) is the injection pumping of electrons
from n-section and holes from p-section in GL and MGL structures with
p-i-n junctions.

In this paper, we study the injection phenomena in GL and MGL p-i-n structures
and calculate their characteristics important for THz lasers.
The idea to use p-n junctions in GLs was put forward and briefly discussed
by us previously~\cite{11}. Here we consider more optimal designs 
of the structures (with a sufficiently  long i-section) and account for
realistic mechanisms of recombination at elevated temperatures (at 
the room temperature and slightly below).

The paper is organized as follows. In Sec.~II, we describe the device structures
under consideration and principles of their operation.
The pertinent equations of the model governing the balance of electrons, holes,
and optical phonons (rate equations) are presented in Sec.~III. 
These equations are reduced to an equation
governing the electron-hole effective temperature. The solution of this equation
in  Secs.~IV and V (both analytically in limiting cases and numerically)
allows us  to find the effective temperature of optical phonons and the current
as functions of the applied voltage, the structural parameters, and the lattice temperature.   In Sec.~VI, the obtained  characteristics of the injected electron-hole plasma
are used to calculate the dependence of the dynamic conductivity
of the latter as a function of the signal THz frequency and other quantities.
Sec.~VII deals with  the model limitations and discussion.
In Sec.~VIII, we draw the main conclusions.

\section{Device model}

\begin{figure}[t]
\center{\includegraphics[width=7.0cm]{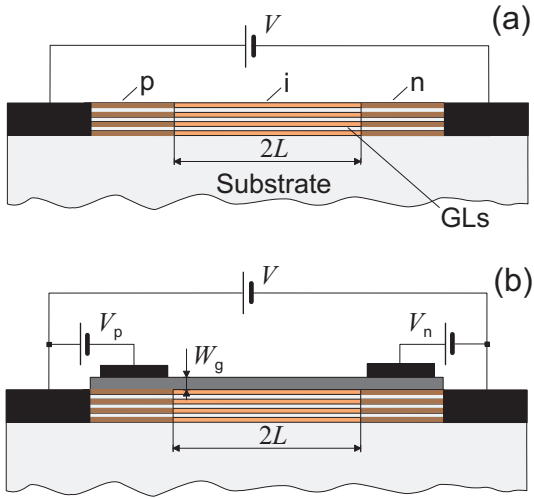}}
\caption{(Color online) Schematic view of the cross-sections
of MGL laser structures  (a) with chemically doped n- and p-sections
and (b) with such section electrically induced by the side gate-voltages
$V_p = -V_g$ and $V_n = V_g >0$. 
}
\label{Fig1}
\end{figure}

\begin{figure}[t]
\center{\includegraphics[width=7.0cm]{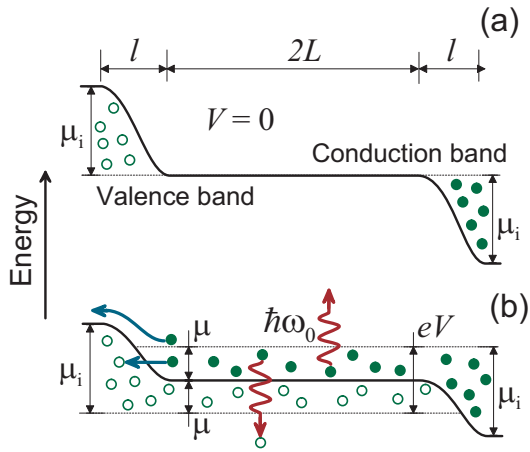}}
\caption{(Color online) Band profiles of a GL in p-i-n junctions
(a) at $V = 0$ and (b) at forward bias $V > 0$.
Opaque and open circles correspond to electrons and holes, respectively.
Wavy arrows show  some  interband (recombination)  and intraband 
transitions  assisted by optical phonons.
Smooth arrows indicate tunneling and thermionic leakage processes.
}
\label{Fig2}
\end{figure}

We consider devices  which comprise   a GL or an MGL structure
with several non-Bernal stacked GLs. It is assumed that
the sections of GLs adjacent to the side contacts are doped
(p- and n-sections). 
The device structure under consideration is shown in Fig.~1(a).
The dc voltage $V$ is applied between 
the side contacts to provide the forward bias of the p-i-n junction. 
Due to doping of the side sections with the acceptor
and donor sheet concentration $\Sigma_i$, the electron and hole 
Fermi energies counted from the Dirac point are $\mu_e = \mu_i$  and
$\mu_h = -\mu_i$, where $\mu_i \simeq \hbar\,v_F\sqrt{\pi\Sigma_i}$,
$\hbar$ is the reduced Planck constant, and $v_F \simeq 10^8$~cm/s
is the characteristic velocity of the carrier spectrum in GLs.
Instead of doping of the side sections, the p- and n-sections can be created using highly conducting gates over these sections to which
the gate voltages $V_p = -V_g < 0$ and $V_n = V_g > 0$ are applied~\cite{11}.
In this case, the chemically doped p- and n-sections are replaced by
the electrically induced sections (see, for instance,~\cite{18,19}) as shown in Fig.~2(b).
In single-GL structures,
 $\mu_i \propto \sqrt{V_g/W_g}$, where $W_g$ is the spacing 
between the GL and the gate.
In the case of MGL structures, the situation becomes more complex due to
the screening of the transverse electric field in GLs~\cite{19}
although the effective electric doping can be achieved in MGL structures with
about dozen GLs.
For definiteness, in the following we shall consider primarily 
the devices with
chemically doped p-i-n junctions.
As shown below, under certain conditions the frequency-dependent dynamic conductivity
of the GL or MGL structures can be negative in a certain range of the signal 
frequencies. In this case,
the self-excitation of THz modes propagating in the substrate serving as a dielectric waveguide (in the direction perpendicular to the injection current) 
and lasing is possible. The metal gates in the devices with electrically induced
p-i and i-n junctions can also serve as the slot-line waveguides for
THz waves. Apart from this, the self-excitation of surface plasmons
(plasmon-polaritons) is possible as well (see, for instance, Refs.~\cite{4,6,8,9,10})

When the p-i-n junction under consideration is forward biased by the applied voltage
$V$,
the electrons and holes are injected to the i-section from the pertinent 
doped side sections. The injected electrons and holes reaching
the opposite doped section can recombine at it due to the interband tunneling
or escape the i-section due to the thermionic processes.
The band profiles in the structures under consideration at $V = 0$
and at the forward bias $V > 0$ are shown in Fig.~2.
Since the probability of such tunneling is a very sharp function
of   angle of incidence, the leakage flux due to the tunneling
electrons (holes) is much smaller that the flux of injected
electrons (holes). The currents associated with the tunneling and thermionic
leakage of
 electrons at the p-i-junction
and holes at the i-n-junction  depend on the electric field at the pertinent
barriers and the applied voltage $V$. Because the relative role of the
leakage currents diminishes with increasing length of the i-section $2L$,
width  of the p-i- and i-n- junctions $l$,   and the barrier height 
 at these junctions $\mu_i$,
we shell neglect it. The pertinent conditions
 will be discussed in the following.  
Thus it is assumed that
the main  fractions of the injected electrons and holes recombine inside
 the i-section.
 The recombination of electrons and holes in GLs  at not too low temperatures is mainly determined by the emission of optical photons~\cite{20}. 
Considering the sub-threshold characteristics (i.e., the states below
the threshold of lasing) and focusing on the relatively high-temperature
operation,
we shall account for this recombination mechanism and disregard
others~\cite{21,22,23,24} including the mechanism~\cite{25} 
associated with
the tunneling between the electron-hole puddles 
(if any)~\cite{26,27,28,29}.
Due to high net electron and hole densities in MGL structures
with sufficient number of GLs,
the latter mechanism can be effectively suppressed~\cite{25}.
We also assume that the net recombination rate in the whole i-section
is much smaller than the fluxes of injected electrons and holes.

\section{Equations of the model}

Due to rather effective inter-carrier scattering, 
the electron and hole distribution functions (at least at not too high energies)
can be very close to
the Fermi distribution functions with quasi-Fermi energies 
$\mu_e$ and $\mu_h$ and the electron-hole
 effective temperature $T$. 
The latter quantities are generally different from those in equilibrium 
(without pumping) at which 
$\mu_e = \mu_h  = 0$ and $T = T_0$.
At the pumping of an intrinsic GL structure, $\mu_e = - \mu_h = \mu$,
where generally  $\mu > 0$.
Under these conditions, the quasi-Fermi energy $\mu$ in the i-section
(neglecting the leakage and recombination in the lowest approximation),
is given by [see Fig.~2(b)]

\begin{equation}\label{eq1}
\mu =  eV/2,
\end{equation}
where $e$ is the electron charge.

The terminal current between the side contacts (per unit length 
in the lateral direction perpendicular to the current),
which coincides with the recombination current, is given by

\begin{equation}\label{eq2}
J =  2eLR_0^{inter}
\end{equation}
The rate of the optical phonon-assisted interband transitions
(recombination rate)  $R_0^{inter}$ and the rate of the intraband energy relaxation associated with optical phonons  $R_0^{intra}$
can be calculated  using the following simplified formulas~\cite{9,16}
(see, also Ref.~\cite{20}:

$$
R_{0}^{inter}
= \frac{\Sigma_0}{\tau_0^{inter}}\biggl[
({\cal N}_0 + 1)\exp\biggl(\frac{2\mu - \hbar\omega_0}{T}\biggr)  
- {\cal N}_0
\biggr] 
$$
\begin{equation}\label{3}
= \frac{\Sigma_0}{\tau_0^{inter}}\biggl[
({\cal N}_0 + 1)\exp\biggl(\frac{eV  - \hbar\omega_0}{T}\biggr)  
- {\cal N}_0
\biggr], 
\end{equation}
\begin{equation}\label{4}
R_0^{intra} = \frac{\Sigma_0}{\tau_0^{intra}}\biggl[
\biggl[
({\cal N}_0 + 1)\exp\biggl(- \frac{\hbar\omega_0}{T}\biggr)  
- {\cal N}_0
\biggr]. 
\end{equation}
Here $\tau_0^{inter}$ and $\tau_0^{intra}$
are  the pertinent 
characteristic times (relatively slow dependent on $\mu$ and $T$), $\Sigma_0$ is the equilibrium electron and hole density,
and ${\cal N}_0$ 
is the number of optical phonons. 
Here and in all equations in the following, $T$ and $T_0$ are in the energy
units. 
When the optical phonon system is close to equilibrium, one can put 
${\cal N}_0 = [\exp(\hbar\omega_0/T_0) - 1]^{-1} = {\cal N}_0^{eq}$.
For numerical estimates we  set   
${\overline R_0}^{inter} = \Sigma_0/\tau_0^{inter} \simeq 10^{23}$~cm$^{-2}$s$^{-1}$ ~\cite{20}. 
Equations~(2) and (3) yield the following general formula for 
the structure current-voltage characteristic:
\begin{equation}\label{5}
J = \frac{2eL\Sigma_0}{\tau_0^{inter}} \biggl[
({\cal N}_0 + 1)\exp\biggl(\frac{eV  - \hbar\omega_0}{T}\biggr)  
- {\cal N}_0
\biggr]. 
\end{equation}
Naturally, at $V = 0$, $T = T_0$, so that ${\cal N}_0 = [\exp(\hbar\omega_0/T_0) - 1]^{-1}$ and $J = 0$.
At $V > 0$, due to contributions of the recombination and injection
to the energy balance of the electro-hole plasma in the i-section,
the electron-hole effective temperature $T$ can deviate from the lattice temperature $T_0$. The number of optical phonons ${\cal N}_0$
can also be different from its equilibrium
value ${\cal N}_0^{eq}$. Since $\hbar\omega_0$ is large, in a wide range of temperatures (including the room temperatures) $\hbar\omega_0 \gg T_0$

The electron-hole plasma gives up the energy $\hbar\omega_0$ in each act
of the  optical phonon emission (interband and intraband) and receives the same 
energy absorbing an optical phonon. Hence,
the  net rate of the energy transfer from and to the electron
hole-plasma due to the inreaction with optical phonons
is equal to  $2L\hbar\omega_0 (R_0^{inter} + R_0^{intra})$.
Considering Eqs.~(3) and (4) and taking into account that
the Joule power associated with the injection current
is equal to
$Q = JV = 2eLR_0^{inter}V$,
an equation governing the energy balance in the electron-hole plasma in the i-section can be presented as

$$
\frac{eV}{\tau_0^{inter}}\,\biggl[
({\cal N}_0 + 1)\exp\biggl(\frac{eV  - \hbar\omega_0}{T}\biggr)  
- {\cal N}_0\biggr]  
$$
$$
= \frac{\hbar\omega_0}{\tau_0^{inter}}
 \biggl[ ({\cal N}_0 + 1)\exp\biggl(\frac{eV  - \hbar\omega_0}{T}\biggr)  
- {\cal N}_0\biggr] 
$$
\begin{equation}\label{6}
+ \frac{\hbar\omega_0}{\tau_0^{intra}}\biggl[
({\cal N}_0 + 1)\exp\biggl(-\frac{ \hbar\omega_0}{T}\biggr)  
- {\cal N}_0\biggr]. 
\end{equation}
Here the left-hand side corresponds to the power received
 by the electron-hole plasma in the i-section 
from the pumping source, whereas
the right-hand side correspond to the power transferred to 
or received from the optical phonon system.

The number of optical phonons is governed by an equation which describes the balance between their generation in the interband and intraband transitions and decay due to the anharmonic contributions to the interatomic potential, leading to the phonon-phonon scattering and in the decay of optical phonons into acoustic phonons. This equation can be presented in the form
$$
\frac{({\cal N}_0 - {\cal N}_0^{eq})}{\tau_0^{decay}} = \frac{1}{\tau_0^{inter}} \biggl[
({\cal N}_0 + 1)\exp\biggl(\frac{eV  - \hbar\omega_0}{T}\biggr)  
- {\cal N}_0\biggr] 
$$
\begin{equation}\label{7}
+ \frac{1}{\tau_0^{intra}}\biggl[
({\cal N}_0 + 1)\exp\biggl(-\frac{ \hbar\omega_0}{T}\biggr)  
- {\cal N}_0\biggr], 
\end{equation}
where 
   $\tau_0^{decay}$ is the optical phonon decay time.
This time can be  markedly longer than $\tau_0^{inter}$ and $\tau_0^{intra}$, 
particularly in suspended GLs, so that parameter 
$\eta_0^{decay} = \tau_0^{decay}/\tau_0^{inter}$ can exceed or substantially exceed
unity.
As shown~\cite{30,31,32,33,34},  
$\tau_0^{decay}$ in GLs
is in the range of  1 - 10~ps. 
As calculated recently~\cite{17}, the charactertic times
$\tau_0^{inter}$ and $\tau_0^{intra}$ can be longer than 1~ps.
If so, the situation when  $\eta_0^{decay} < 1$ appears also to be feasible.
The optical phonon decay time might be fairly short depending on the type of the substrate.

Instead of Eq.~(7) one can use the following equation
which explicitly reflexes the fact that the energy received by the electron-hole plasma from the external voltage source goes eventually to
the optical phonon system: 

\begin{equation}\label{8}
\eta_0^{decay}\frac{eV}{\hbar\omega_0}\,\biggl[
({\cal N}_0 + 1)\exp\biggl(\frac{eV  - \hbar\omega_0}{T}\biggr)  
- {\cal N}_0\biggr]  =  {\cal N}_0 - {\cal N}_0^{eq}.
\end{equation}
Using Eqs.~(6) and (7) or Eqs.~(6) and (8), one can find $T$ and ${\cal N}_0$ as functions of 
$V$ and then calculate the current-voltage characteristic
invoking Eq.~(5), as well as the dynamic  characteristics.

Equation~(8) yields
\begin{equation}\label{9}
{\cal N}_0 = 
\frac{{\cal N}_0^{eq} + \displaystyle\eta_0^{decay}\frac{eV}{\hbar\omega_0}\exp\biggl(\frac{eV - \hbar\omega_0}{T}\biggr)}{1 + \displaystyle\eta_0^{decay}\frac{eV}{\hbar\omega_0}\biggl[1 - \exp\biggl(\frac{eV - \hbar\omega_0}{T}\biggr)\biggr]}.
\end{equation}
Substituting ${\cal N}_0$ given by Eq.~(9) to Eq.~(6), we arrive at the following equation for $T$:
$$
\frac{1 + 
\displaystyle\eta_0^{decay}\frac{eV}{\hbar\omega_0}\biggl[1 - 
\exp\biggl(\frac{eV - \hbar\omega_0}{T}\biggr)\biggr]}
{{\cal N}_0^{eq} + \displaystyle\eta_0^{decay}\frac{eV}{\hbar\omega_0}\exp\biggl(\frac{eV - \hbar\omega_0}{T}\biggr)}
$$
$$
\times\biggl[\eta_0\frac{(\hbar\omega_0 - eV)}{\hbar\omega_0}
\exp\biggl(\frac{eV}{T}\biggr) + 1\biggr]
$$
$$
+ \eta_0\frac{(\hbar\omega_0 - eV)}{\hbar\omega_0}
\biggl[\exp\biggl(\frac{eV -\hbar\omega_0}{T}\biggr) - 1\biggr]
\exp\biggl(\frac{\hbar\omega_0}{T}\biggr)
$$ 
\begin{equation}\label{10}
- 
\exp\biggl(\frac{\hbar\omega_0}{T}\biggr) +1 = 0.
\end{equation}
The ratio 
$\eta_0 = \tau_0^{intra}/\tau_0^{inter}$  is actually a function 
of $\mu$ and $T$. The $\eta_0 - \mu$ and $\eta_0 - T$ dependences
are associated with the linearity of the density of states in GLs 
as a function of energy. 
To a good approximation these dependences  can be described~\cite{16} 
 by  function
$\eta_0 = \hbar^2\omega_0^2/(6\mu^2 +
\pi^2T^2)$ with $\eta_0 \propto (\hbar\omega_0/T)^2$
at $\mu \ll T$ and $\eta_0 \propto (\hbar\omega_0/\mu)^2$
at $\hbar\omega_0 > \mu \gg T$. 
Thus, considering Eq.~(1), $\eta_0$ in Eq.~(10) is given by
\begin{equation}\label{11}
 \eta_0 = \frac{2\hbar^2\omega_0^2}{(3e^2V^2 +
2\pi^2T^2)}.
\end{equation}

Introducing the effective temperature of the optical phonon system 
$\Theta$ such that
${\cal N}_0 = [\exp(\hbar\omega_0/\Theta)-1]^{-1}$, i.e., 

\begin{equation}\label{12}
\Theta = \frac{\hbar\omega_0}{\ln (1 + {\cal N}_0^{-1})},
\end{equation}
and substituting ${\cal N}_0$ from Eq.~(9) to Eq.~(11), 
one can relate $\Theta$ and $T$. Then calculating the $T - V$ dependences
using Eq.~(10), one can find the pertinent $\Theta - V$ dependences.

\section{Effective temperatures and
current-voltage characteristics (analytical analysis)}
\subsection{Low voltages} 
In particular, at $V = 0$, Eqs.~(8) and (10)  naturally  yield
${\cal N}_0 = {\cal N}_0^{eq}$ and $T = T_0$. 
At sufficiently low voltages when  $ (\eta_0^{decay}eV/\hbar\omega_0) \ll 1$, 
the solutions of Eqs.~(9) and (10) can be found
 analytically. 
In this case, ${\cal N}_0 \simeq {\cal N}_0^{eq}$.
Considering this,
at low voltages, Eq.~(10)  yields

\begin{equation}\label{13}
T \simeq T_0 \biggl[1 - \frac{eV}{\hbar\omega_0}\frac{\eta_0^{eq}}{(1 + \eta_0^{eq})}\biggr] < T_0,
\end{equation}
where $\eta_0^{eq}$ is the value of 
$\eta_0$ at $V \ll \hbar\omega_0/e$~,
i.e., $\eta_0^{eq} \simeq 5$~\cite{16}.
%
%
As follows from Eq.~(13), an increase in the applied voltage $V$ leads
to a decrease in the effective temperature 
of the electron-hole plasma (its cooling). 

Using Eqs.~(5) and (13) at low voltages, we also obtain
\begin{equation}\label{14}
\frac{J}{J_0} \simeq \frac{eV}{T_0}\frac{\eta_0^{eq}}{(1 + \eta_0^{eq})} 
\exp\biggl(-\frac{\hbar\omega_0}{T_0}\biggr).
\end{equation}
Here  the current $J$ is normalized
by its characteristic value $J_0 = 2eL\Sigma_0/\tau_0^{inter}
= 2eL{\overline R^{inter}}$.

\subsection{Special cases}

In the special case  $V = \hbar\omega_0/e$ (i.e., $V \simeq 0.2$~V), 
from Eqs.~(9), (10), and  (12)
we obtain 
\begin{equation}\label{15}
T = \frac{\hbar\omega_0}
{\ln\biggl(\displaystyle\frac{1 + \eta_0^{decay} + {\cal N}_0^{eq}}
{\eta_0^{decay} + {\cal N}_0^{eq}}
\biggr)},
\end{equation}
\begin{equation}\label{16}
{\cal N}_0  = {\cal N}_0^{eq} + \eta_0^{decay}.
\end{equation}
One can see that in this  case 
\begin{equation}\label{17}
{\cal N}_0 = \frac{1}{\exp(\hbar\omega_0/T) - 1},
\end{equation}
i.e., $\Theta = T$ and
\begin{equation}\label{18}
J = 
J_0.
\end{equation}
At $T_0 = 300$~K
and $2L = 20~\mu$m,
one obtains  $J_0 \simeq 32$~A/cm. 
As follows from Eqs.~(15) - (17)
at $V = \hbar\omega_0/e$,  $T$ and $\Theta$ tend to $T_0$
if $\eta_0^{decay}$ tends to zero, and
$T = \Theta$, and they both increase proportionally to  $ \eta_0^{decay}$
as  $\eta_0^{decay}$ tends to infinity. 
 Thus, at $V = \hbar\omega_0/e$ and 
$\eta_0^{decay} \sim 1$, the effective temperatures are fairly high: 
$T = \Theta \sim \hbar\omega_0/\ln 2$ ($T = \Theta \sim 3300$~K).
It is worth noting  that $J$ at $V = \hbar\omega_0/e$ is independent of parameter
$\eta_0^{decay}$. 

In interesting (but nonrealistic) limiting  case $\eta_0^{decay} = 0$,
from Eq.~(8) 
we immediately obtain ${\cal N}_0 = {\cal N}_0^{eq}$.
In such a case, $T = T_0$ both at $V = 0$ and $V =\hbar\omega_0/e$.
If $V = \hbar\omega_0/2e$, one obtains
\begin{equation}\label{19}
T \simeq \frac{T_0}{2[1 - (T_0/\hbar\omega_0)\ln (2 + \eta_0)]},
\end{equation}

\begin{equation}\label{20}
J \simeq J_0(\eta_0 + 1)\exp\biggl(- \frac{\hbar\omega_0}{T_0}\biggr) \ll J_0.
\end{equation}
At $T_0 = 300$~K , Eqs.~(19) and (20)  yield,
$T \simeq 0.60T_0 =  180$~K and
$J \simeq 31$~mA/cm at $T_0 = 300$~K
and 
$T \simeq 0.569 T_0 = 114$~K and $0.64$~mA/cm at $T_0 = 200$~K.

\subsection{Long optical phonon decay time}

In  the case of relatively long optical decay time when
 $\eta_0^{decay} \gg 1$ at $V \lesssim \hbar\omega_0/e$,
 neglecting terms of the order of ${\cal N}_0^{eq}/\eta_0^{decay}
\simeq\exp(-\hbar\omega_0/T_0)/\eta_0^{decay}$,
from Eq.~(10) we obtain

\begin{equation}\label{21}
T \simeq \frac{eV}
{\ln\biggl\{ \displaystyle \frac{ 1 +\eta_0^{decay}(eV/\hbar\omega_0) }
{\displaystyle\eta_0^{decay}(eV/\hbar\omega_0) + 
\eta_0[(eV/\hbar\omega_0) - 1]}\biggr\}}.
\end{equation}
At $V = \hbar\omega_0/e$, 
Eq.~(21) yields the same value of $T$ as
Eq.~(15) provided $\eta_0^{decay} \gg {\cal N}_0^{eq}$.


\section{Effective temperatures and current-voltage characteristics
(numerical results)}
\begin{figure}[t]
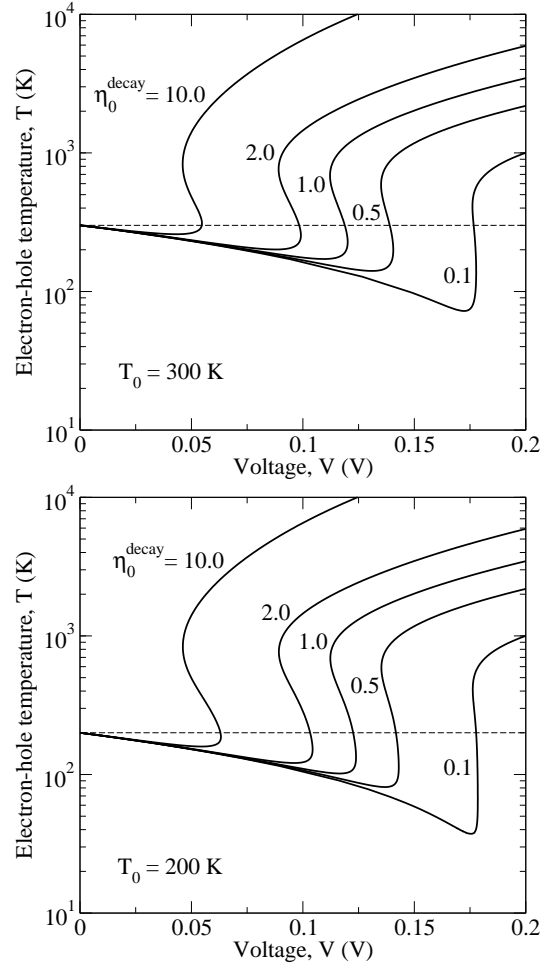
\label{F3AB}
\center{
\includegraphics[width=7.0cm]{INJLASER_F3A.eps}\\
\includegraphics[width=7.0cm]{INJLASER_F3B.eps}}
\caption{Electron-hole effective temperature $T$ as 
function of applied voltage $V$ for different values of $\eta_0^{decay}$
and $T_0$.
Dashed lines correspond to $T = 300$K (upper panel) and $T = 200$~K (lower panel).
}
\end{figure}
\begin{figure}[t]\label{F4}
\center{
\includegraphics[width=6.8cm]{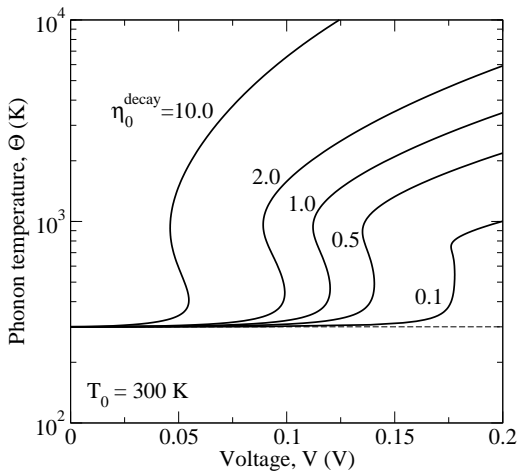}
\caption{Optical phonon  effective temperature $\Theta$ as 
function of applied voltage $V$ for different values of $\eta_0^{decay}$. 
Dashed line corresponds to $\Theta = 300$~K.
}
}
\end{figure}

To obtain $T -V$ and $\Theta - V$ dependences in wide ranges of 
parameter $\eta_0^{decay}$ and the applied voltage $V$,
Eqs.~(9) - (12) were  solved numerically.
Figures 3 and 4  show the voltage dependences of  effective
temperatures $T$ and $\Theta$
calculated for different values 
$\eta_0^{decay}$ and $T_0 = 300$ and 200~K. 
One can see from Fig.~3 that the electron-hole effective temperature
markedly decreases with increasing voltage, so that $T < T_0$ 
in a certain voltage range [see also Eq.~(13)],
 and then starts to rise. 
In the range of relatively high voltages ($V \lesssim \hbar\omega_0/e$),
the $T - V$ dependence is steeply rising
[in line with analytical formula given by  Eq.~(21)] 
with $T > T_0$ or even $T \gg T_0$.
However it is intriguing that in a rather narrow voltage range
where $V$ is about some value $V_d$, the $T - V$ dependences are ambiguous, so that
these dependences as a whole are of the S-shape.
The appearance of the S-shape characteristics can be attributed to
a decrease in parameter $\eta_0$ with increasing $T$ [see Eq.~(11)]. 
This corresponds to a decrease in $\tau_0^{intra}$, and, hence, to
an essential intensification of the intraband transitions, particularly,
those associated with the reabsorption of nonequilibrium optical phonons
when $T$ increases. This is because at high electron-hole effective  temperatures
the intraband transitions assisted by optical phonons take place between
relatively high energy states with their elevated density.
When $V$ exceeds some ``disruption'' voltage $V_d$, the net power acquired by
the electron-hole plasma can be compensated by the intraband energy relaxation on optical phonons only at sufficiently high $T$. As a result, in this case
the electron-hole temperature jumps to the values corresponding to higher
branch of the $T - V$ dependence. Thus, the ``observable'' $T - V$
dependences and their consequences can as usual exhibit  hysteresis
instead of the S-behavior.
One needs to point out that if the above temperature-dependent parameter $\eta_0$
is replaced in calculations by a constant, the calculated $T - V$ dependences
become unambiguous, although they exhibit a steep increase in the range 
$V \sim V_{d}$.

The behavior of $T$ as a function of $V$ markedly depends on parameters
$\eta_0^{decay}$ and $\eta_0$.
The width of the voltage  range where $T < T_0$ increases 
when parameter $\eta_0^{decay}$ becomes smaller
with increasing voltage.
Simultaneously, the depth of the $T - V$ dependence sag with $T < T_0$
increases with decreasing $\eta_0^{decay}$ as well as 
 with decreasing $T_0$.
At small $\eta_0^{delay}$, the electron-hole cooling can be rather
strong, particularly when $T_0 = 200$~K. 
This is natural because faster decay of optical phonons prevents their
accumulation (heating)  and promotes the electron-hole plasma cooling  when
the Joule power is smaller than the power transferred from electrons and holes to
optical phonons. It worth noting that the voltage range where the $T - V$
dependence is ambiguous widens with increasing $\eta_0^{decay}$.  


As seen from Fig.~4, the optical phonon effective temperature also exhibits
 a $S$-shape voltage dependence. However, contrary to the electron-hole effective temperature, $\Theta \geq T_0$ at all the voltages under consideration.
The values of $\Theta$ at relatively high voltages steeply increase with increasing parameter $\eta_0^{decay}$.


Comparing the $T - V$ dependences calculated for different lattice temperatures, one can find that
at moderate and large
values of  $\eta_0^{delay}$, these dependences are 
virtually independent of $T_0$. This is because in such a case 
the number of optical phonons
${\cal N}_0 \gg {\cal N}_0^{eq}$ and, hence, $\Theta \gg T_0$ even
at not too high voltages, so that the role of equilibrium optical phonons 
is weak.

Invoking Eq.~(5), the $T -V$ dependences  obtained above
can be used to find the current-voltage characteristics.
Figure~5 shows the $J - V$ characteristics calculated
using Eq.~(5) and the $T - V$ and $\Theta - V$ dependences
obtained numerically.
As a consequence of the S-shape $T - V$ and $\Theta - V$ dependences,
the $J - V$ characteristics (as well as the voltage dependences of
the dynamic conductivity considered in the following)  are also of the $S$-shape.
According to Figs.~3 and 5, the $T - V$ and $J - V$ characteristics
in the range of low and moderate voltages 
are independent of parameter $\eta_0^{decay}$. This is in line
with the results
of the previous analytical analysis [see Eqs.~(13) and (14)]. 
However, at relatively high voltages, distinctions in the $J - V$ characteristics for different $\eta_0^{decay}$ is significant although all of them tend to $J_0$
when $V$ approaches to $\hbar\omega_0$/e.

\begin{figure}[t]\label{F5}
\center{
\includegraphics[width=6.8cm]{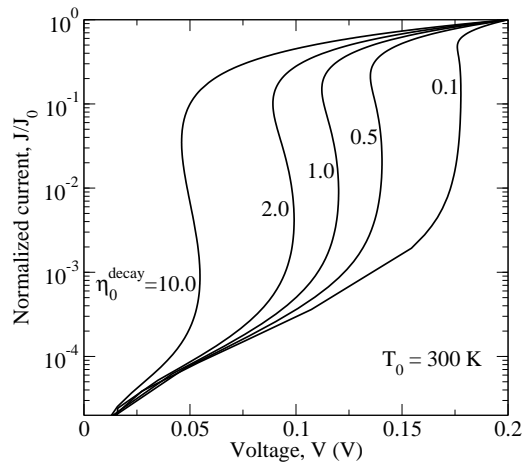}\\
\caption{Normalized current-voltage characteristics for  for different values of $\eta_0^{decay}$.}
}
\end{figure}

\begin{figure}[t]
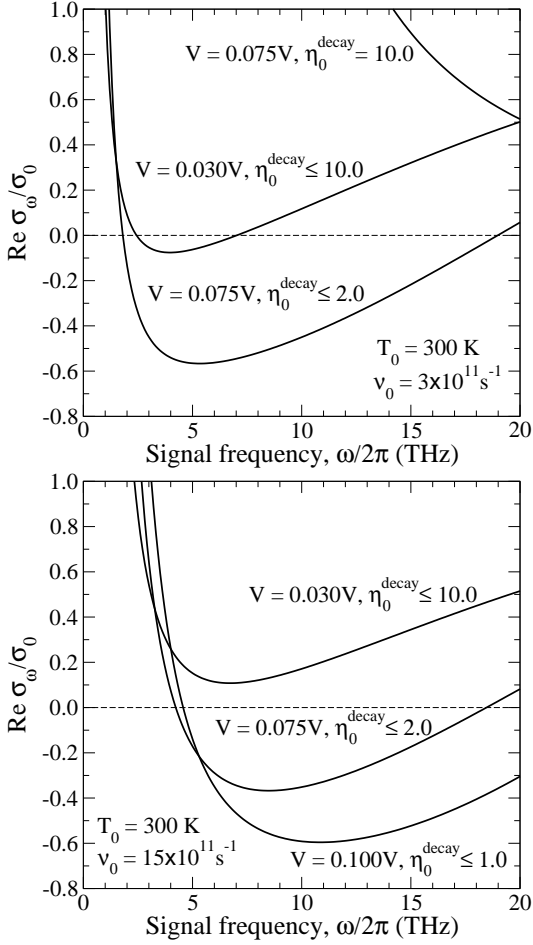
\label{F6}
\center{
\includegraphics[width=7.0cm]{INJLASER_F6A.eps}\\
\includegraphics[width=7.0cm]{INJLASER_F6B.eps}
\caption{Dependence of normalized dynamic conductivity  
Re~$\sigma_{\omega}/\sigma_0$
on signal frequency
$\omega/2\pi$
 for   different values of $\eta_0^{decay}$
and  $\nu_0 \simeq 3\times 10^{11}$~s$^{-1}$ (upper panel)
and  $\nu_0 \simeq 15\times 10^{11}$~s$^{-1}$ (lower panel).
}
}
\end{figure}

\begin{figure}[t]
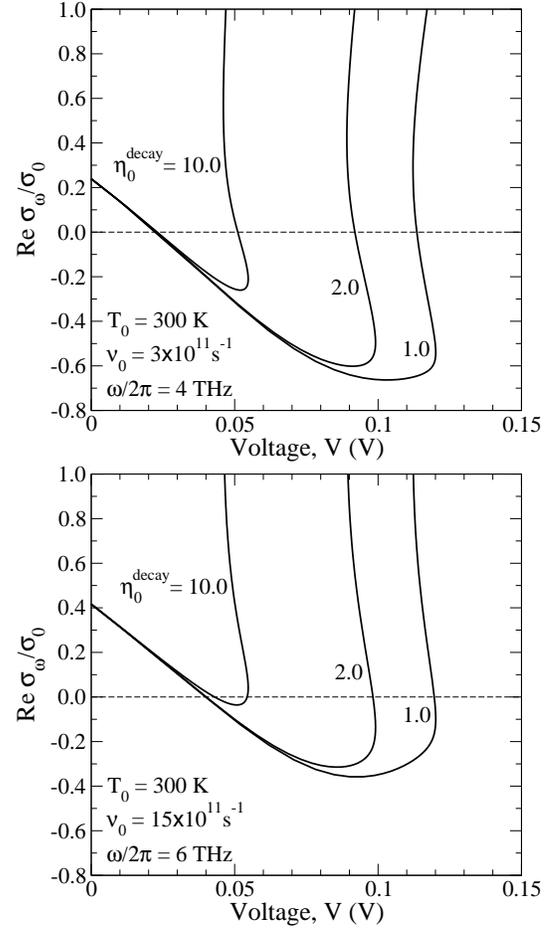
\label{F7}
\center{
\includegraphics[width=7.0cm]{INJLASER_F7A.eps}\\
\includegraphics[width=7.0cm]{INJLASER_F7B.eps}
\caption{Normalized dynamic conductivity Re~$\sigma_{\omega}/\sigma_0$
 vs applied voltage $V$
 for   different values of $\eta_0^{decay}$
and  $\nu_0 \simeq 3\times 10^{11}$~s$^{-1}$, $\omega/2\pi = 4$~THz (upper panel)
and $\nu_0 \simeq 15\times 10^{11}$~s$^{-1}$,  
$\omega/2\pi = 6$~THz (lower panel).
}
}
\end{figure}

\section{Dynamic conductivity}
Knowing the $T - V$ dependences, one can 
calculate the dynamic conductivity,  $\sigma_{\omega}$,  of a GL
under the injection pumping
as a function of the signal frequency $\omega$ and the applied voltage
$V$. To achieve lasing at the frequency $\omega$, 
the real part of the complex
dynamic conductivity at this frequency should be negative:  
Re~$\sigma_{\omega} < 0$.
As shown previously (see, for instance, Refs. ~\cite{4,35}),
the interband contribution of the nonequilibrium electron-hole plasma
with the quasi-Fermi energy $\mu$ and the effective temperature
$T$ is proportional to $\tanh[(\hbar\omega - 2\mu)/4T]$.
The intraband contribution to  Re~$\sigma_{\omega}$, which corresponds to the Drude absorption, depends on $\omega$, $\mu$, and  $T$ as well.
It also depends on the time of  electron and hole 
momentum relaxation on impurities and phonons $\tau$. 
The latter is a function of the energy of electrons 
and holes $\varepsilon$.
The main reason for the  $\tau - \varepsilon$ dependence is a linear
increase in the density of state in GLs with increasing $\varepsilon$.
In this case, $\tau^{-1} = \nu_0(\varepsilon/T_0)$, where $\nu_0$
is the collision frequency of electrons and holes in equilibrium at $T = T_0$. Considering this and taking into account Eq.~(1),
we can arrive at the following  formula~\cite{16} approximately valid in the frequency range $\omega \gg \nu_0$:

\begin{equation}\label{22}
\frac{{\rm Re}\sigma_{\omega}}{\sigma_0} = 
\tanh\biggl(\frac{\hbar\omega - eV}{4T}\biggr)
+
C\frac{(e^2V^2 + 2\pi^2T^2/3)}{\hbar^2\omega^2}.
\end{equation}
Here $\sigma_0 = e^2/4\hbar$ and
$C = 2\hbar\nu_0/\pi T_0$.
In high quality MGLs (with $\nu_0^{-1} \simeq 20$~ps at $T_0 =50$~K~\cite{2,3}),
assuming that $\nu_0 \propto T_0$~\cite{36}, for $T_0 = 300$~K
one can set $\nu_0 \simeq 3\times 10^{11}$~s$^{-1}$ and hence,
$C \simeq 0.005$. For  substantially less perfect GLs with 
$\nu_0 \simeq 15\times 10^{11}$~s$^{-1}$, one obtains $C \simeq 0.025$.
These data are used for the calculations of  Re~$\sigma_{\omega}$.

Figure~6 demonstrates the frequency dependences of  
Re~$\sigma_{\omega}/\sigma_0$ calculated using Eq.~(22) for different values of parameter
$\eta_0^{decay}$ at different voltages $V$
for
  $\nu_0 \simeq 3\times 10^{11}$~s$^{-1}$
and $\nu_0 \simeq 15\times 10^{11}$~s$^{-1}$.
As seen from Fig.~6
at the injection conditions under consideration,
the characteristic conductivity $\sigma_0$ is much smaller than
the dc conductivity in the i-section $\sigma_{\omega}|_{\omega = 0} = 
\sigma_{00}$. It is also seen that even at relatively large values of
 parameter
$\eta_0^{decay}$, the dynamic conductivity can be  negative in the THz range of frequencies provided the applied voltage is not so strong to cause the electron-hole plasma and optical phonon system overheating and the ambiguity of the voltage
characteristics. This is confirmed by Fig.~7.
Figure~7 shows   Re~$\sigma_{\omega}/\sigma_0$ as a function of 
the applied voltage.
As demonstrated, the range of the signal frequencies
where Re~$\sigma_{\omega} < 0$ markedly shrinks and  the quatity
$|{\rm Re}~\sigma_{\omega}|$ decreases when either $\eta_0^{decay}$
or $\nu_0$ increase. In particular, at large values of $\eta_0^{decay}$,
the achievement of the negative dynamic conductivity and THz lasing
can be complicated  (at the temperatures $T_0 \sim 300$~K when the optical phonon recombination mechanism dominates). This is because when 
$\eta_0^{decay}$ increases,  the quantity $V_d$ becomes small. As a result,  
the optical phonon system 
 is overheated starting from relatively low  voltages  
that leads
to an ``early''  overheating of the electron-hole plasma 
[see Figs.~(3) and (4)].
If the value of Re~$\sigma_{\omega}$ is insufficient to overcome the 
losses of the THz modes propagating along the GL structure, 
the structures with MGL can be used. In this case, the net dynamic conductivity
of the MGL structure is given by Re~$\sigma_{\omega}\times K$, where $K$ is the number of GLs~\cite{8,9,10}.

\section{Limitations of the model and discussion}

High density of the electron-hole plasma in the i-section under the injection
conditions promotes the quasi-neutrality of this section.
The recombination does not significantly affect 
the uniform  distribution of the  electron and hole densities 
in the i-section (assumed above)
if the recombination current $J$ is much smaller than
the maximum injection current $J_m$ which can be provided by
the p- and n-sections. This imposes the condition 
$J \ll 2J_m$. The latter inequality is equivalent 
to the condition that
the recombination length is  longer than the i-section length $2L$.
Considering a strong degeneracy of the electron and hole components
in the p- and  n-sections, respectively,
the quantity $J_m$ can be estimated as
\begin{equation}\label{23}
J_m \simeq \frac{ev_F}{\pi^2\hbar^2}[\mu_i^2 - (\mu_i - eV)^2].
\end{equation}
At $T_0 = 300$~K, $2L = 20~\mu$m, $\mu_i = 0.3$~eV, and $V = 0.1$~V, 
one obtains $J_0 \simeq 32$~A/cm and
$2J_m \simeq 65$~A/cm.
As seen from Fig.~5,  $J \ll J_0$ at least 
in the most interesting voltage range, where Re $\sigma_{\omega} < 0$
(see Fig.~7), the condition $J \ll J_{m}$ is satisfied.

In the above consideration we disregarded 
the leakage current from the i-section to the p-
and n-sections. This current  includes the tunneling and thermionic components.
Both these components depend on the height, 
$\mu_i$,  of the barriers between
the p- and i-sections and i- and n-sections and the applied voltage $V$
(see Fig.~2).
As shown previously~\cite{37}, the tunneling current decreases with 
increasing width of the p-i- or i-n junction $l$ [see Fig.~2(a)], 
because it is sensitive to the electric field 
at the junction~\cite{18}. 
The width in question depends on the geometrical parameters 
of the structure, in particular on the spatial distributions 
of donor and acceptors
near the junction,  and the thickness of the gate layer $W_g$, as well as
 the shape
of the gates (in the structures with the electrical doping).
So one can assume that this width can be sufficiently large to provide smooth
potential distributions at the junctions.
The effective height of the barriers at the p-i- and i-n-junctions, which determines the thermionic electron and hole current over these barriers,
is equal to $\Delta = \mu_i - 2\mu = \mu_i - eV$. It can be small  
at elevated values of the quasi-Fermi energy $\mu$ which are necessary to achieve
the negative dynamic conductivity (see, for instance, Refs.~\cite{4,6,9,10}),
i.e., in the most interesting case. This, in turn, implies 
that the electric field at the junctions and, hence, the tunneling current
is decreased. 
In this case, the thermionic leakage current dominates over 
the tunneling leakage current, and the latter is  disregarded in the following estimates.

Taking into account  the height of the barrier $\Delta$, for 
the  contribution of thermionic current through the junctions to the net terminal current one can obtain
$$
J_{th} = \frac{4ev_F}{\pi^2}\biggl(\frac{T^2}{\hbar^2v_F^2}\biggr)^2
\biggl[\exp\biggl(\frac{eV - \mu_i}{T}\biggr) - 1\biggr]
$$
\begin{equation}\label{24}
= 2ev_{th}\Sigma_0\biggl(\frac{T}{T_0}\biggr)^2\biggl[\exp\biggl(\frac{eV - \mu_i}{T}\biggr) - 1\biggr].
\end{equation}
Here
$v_{th} = (12/\pi^3)v_F$.
Comparing $J$ and $J_{th}$ given by Eqs.~(5) and (24), respectively,
one can conclude that the  leakage (thermionic) current is small in comparison with the current associated with the optical phonon recombination if
$(v_{th}\tau_0/L)\exp(- \mu_i/T) \ll \exp(- \hbar\omega_0/T)$ or
\begin{equation}\label{25}
\mu_i - \hbar\omega_0 >  T\ln \biggl[\biggl(\frac{v_{th}\tau_0}{L}\biggr)
\biggl(\frac{T}{T_0}\biggr)^2\biggr].
\end{equation}
Assuming that $\tau_0 \sim 10^{-12}$~s, factor $(v_{th}\tau_0/L)$ in Eq.~(21)
is small when the length of the i-section $2L \gtrsim 1~\mu$m, 
i.e., at fairly practical values of $2L$.
Therefore, neglect of the thermionic leakage in our calculations in
the above sections is justified when $\mu_i > \hbar\omega_0 \simeq 0.2$~eV
(more precisely when $\mu_i - \hbar\omega_0 > T$).
At $T \gtrsim T_0$,
the latter inequality means that the donor and acceptor density
in the pertinent sections should be
\begin{equation}\label{26}
\Sigma_i > \frac{1}{\pi}\biggl(\frac{\omega_0}{v_F}\biggr)^2
\simeq 3.3\times10^{12} {\rm cm}^{-2}.
\end{equation}

In the case of the devices with the electrically-induced p- and n-section,
the analogous condition sounds as

\begin{equation}\label{27}
V_g > \frac{4eW_g}{\ae}\biggl(\frac{\omega_0}{v_F}\biggr)^2.
\end{equation}
Setting $\ae = 4$ and $W_g = 10$~nm, the latter condition corresponds to
$V_g > 1.5$~V.

At a strong heating of the electron-hole plasma, say, at $V = \hbar\omega_0/e$,
using Eq.~(15), condition~(25)
is replaced by

\begin{equation}\label{28}
\mu_i > \hbar\omega_0\frac{\ln(v_{th}\tau_0/L)}
{\ln\biggl(\displaystyle\frac{1 + \eta_0^{decay} + {\cal N}_0^{eq}}
{\eta_0^{decay} + {\cal N}_0^{eq}}
\biggr)}.
\end{equation}
The latter inequality can impose somewhat stricter limitation 
on the values of $\mu_i$ 
and $\Sigma_i$
than
those given by Eqs.~(25) and (26) if $\eta_0^{decay} > 1$. 

It is notable that an effective confinement of the injected electrons 
and holes in the i-section by the barriers at p-i- and i-n-junctions
in the GL structures under consideration can be realized by relatively
low doping levels in p- and n-sections in comparison with
structures with two-dimensional (2D) electron plasma 
in quantum wells on the base of
the standard semiconductors.
Indeed, the barrier height at $V = 0$ in the doped section of GL is equal to 
$\Delta = \mu_i -  eV$. In standard 2D   systems
with the electron effective mass $m^*$ and the energy gap $\Delta_g$, 
the barrier height is equal to $\Delta^{*} = \Delta_g + \mu_i^{*} - eV^{*}$,
where $\mu_i^{*} \simeq \pi\hbar^2\Sigma_i^{*}/m^*$.
To achieve the same value of the quasi-Fermi energy $\mu$
in the standard 2D system as in GLs, one needs to apply the voltage
$V^{*} = V + \Delta_g/e$.
 To provide, for example,  the values $\mu_i = \mu_i^* = 0.3$~eV, one needs
$\Sigma_i \simeq 7.33\times 10^{12}$~cm$^{-2}$ 
and  $\Sigma_i^* \simeq 3.82\times 10^{13}$~cm$^{-2}$, respectively.
Thus, in the standard 2D electron systems the doping should be 
 five times higher  (for $m = 4\times 10^{-29}$~g).  
This is due to lower density of states in massless GLs near the Dirac point
compared to that in the standard 2D structures with $m \neq 0$.
Since the effective mass of holes $M^*$ in the standard
semiconductors is markedly larger than $m^*$, the realization of the barrier
height at the p-i-junction sufficient for the effective confinement of electrons at elevated temperatures requires fairly heavy doping.
Thus, in contrast to  the standard 2D structures,   the thermionic
leakage current in the GL or MGL  p-i-n structures under consideration
  can be sufficiently small without the employment of wide-gap p- and n-sections.
In passing  it should be mentioned that an extra confinement of the injected
electrons and holes can be achieved if the p- and n-sections constitute
arrays of graphene nanowires (doped or with electrically-induced high electron an hole densities), so that the p-i-n structures considered above are replaced
by the P-i-N structures.

One needs to stress that the assumptions (used in the above model)
that the electron-hole plasma in
the active region 
 is virtually uniform as well as
 that the recombination current exceeds the leakage current
are rather common in simplified models (the so-called rate-equation models) of 
in standard injection laser structures  with the double injection
(see, for instance, Refs.~\cite{38,39}).

As follows from the above calculations,
the effective electron-hole and optical phonon temperature
can be very high at $V \sim \hbar\omega_0/e$ and be accompanied by effects associated with the $S$-shape characteristics. 
In this case, an expression for the rate of optical phonon decay 
 $({\cal N}_0 - {\cal N}_0^{eq})/\tau_0^{decay}$ used in Eq.~(7)
might  be oversimplified
due to a strong anharmonism of the lattice vibration.
Possibly, the effects related to  a strong anharmonism  can be taken into account by a proper choice (renormalization) of parameters $\tau_0^{decay}$ and
 $\eta_0^{decay}$. In the above treatment, for simplicity only one type of optical phonons with $\hbar\omega_0 \simeq 0.2$~eV was taken in to account.
However, due to closeness of the optical phonon frequencies of different type
in GLs and MGLs, the pertinent generalization of the model, adding computational
complexity,
 should not lead
to a marked change in the obtained results. 
Apart from this, at large effective temperatures, the radiative recombination
and  cooling
(due to the radiative transfer of the energy outside the structure)
can become essential~\cite{40,41,42}, resulting in a limitation  
of these temperatures and affecting the $S$-shape dependences. 
This means that considering
the range or relatively high applied voltages and, hence, strong injection,
 our purely ``optical phonon'' model should be generalized.
However, this  concerns not particularly interesting
 situations in which the dynamic conductivity is not negative.


As demonstrated, the main potential obstacles in the realization
of negative dynamic conductivity and THz lasing in the injection GL and MGL structures at the room (or slightly lower) 
temperatures might be  the intraband photon (Drude) absorption
and the optical phonon heating. These  effects are characterized
by parameters $C \propto \nu_0$ and $\eta_0^{decay} \propto \tau_0^{decay}$,
respectively. As for parameter $C$, it can be sufficiently small in perfect
MGL structures like those studied in Ref.~\cite{2}, so the problem of 
intraband absorption can be overcome.
However, if real values of  parameter  $\eta_0^{decay}$ can not be decreased
to an appropriate level ($\eta_0^{decay} \lesssim 1$), 
the achievement of room temperature THz lasing
in the structures under consideration might meet problems.
In the case of such a scenario, the utilization of lower temperatures, at which
the recombination and energy relaxation is associated with different
mechanisms,
can become indispensable.

\section{Conclusions}

In conclusion, we have
studied theoretically
 the effect of population inversion associated with the electron and hole injection in GL and MGL  p-i-n structures at the room and slightly lower
temperatures when the interaction with optical phonons
is the main mechanism of the recombination and energy relaxation.
In the framework of the developed model,
 the electron-hole and optical phonon  effective temperatures and
the current-voltage characteristics
 have been 
calculated as functions of the applied voltage and the structure parameters.
It has been demonstrated  that the injection can lead to cooling of the injected electron-hole plasma in the device i-section to the temperatures lower
than the lattice temperature at low and moderate voltages, whereas 
the voltage dependences can be ambiguous  exhibiting the S-shape
behavior at elevated voltages.
The  variations of the electron-hole effective temperature with increasing
applied voltage are accompanied with an increase in the optical phonon effective
temperature.
Using the obtained voltage  dependences, we have calculated the  dynamic conductivity
and estimated  the ranges parameters and signal THz frequencies
where this conductivity is negative.
The electron-hole cooling  might substantially promote 
the realization of THz lasing at elevated ambient temperatures. 
In summary, we believe that the obtained results instill confidence
in the future of graphene-based injection  THz lasers
although their realization might require a thorough 
optimization.

\section*{Acknowledgment}

The authors are grateful to A.~Satou for numerous useful discussions.
This work was supported by the Japan Science and Technology Agency, CREST and by the 
Japan Society
for Promotion of Science, Japan.

\end{document}